\documentstyle[12pt,fleqn,epsf0]{article}
\def\beeq{\begin{equation}}
\def\eneq{\end{equation}}

\addtocounter{section}{-1}
\begin{document}

\begin{center}

\vspace{2cm}

{\large {\bf {
Persistence of Edge-State in Stacked Graphene\\
and Nano-Graphene Materials
} } }

\vspace{1cm}

{\rm Kikuo Harigaya\footnote[1]{E-mail address: 
\verb+k.harigaya@aist.go.jp+; URL: 
\verb+http://staff.aist.go.jp/k.harigaya/+}}

\vspace{1cm}

{\sl Nanotechnology Research Institute, AIST, 
Tsukuba 305-8568, Japan}\footnote[2]{Corresponding address}\\

\vspace{1cm}

(Received~~~~~~~~~~~~~~~~~~~~~~~~~~~~~~~~~~~)
\end{center}

\vspace{1cm}

\noindent
{\bf Abstract}\\
Nano-carbon materials are investigated intensively. In this paper,
the edge-state in nanographene materials with zigzag edges is 
studied theoretically.  In particular, while the inter-layer 
interactions are considered, we prove that edge states 
exist at the energy of the Dirac point in the doubly 
stacked nanographene, and in the case of the infinitely-wide
lower layer case.  This property applies both for the A-B 
and A-C stackings.

\pagebreak

\section{Introduction}

The graphite and single layer graphene materials have
been studied intensively, since the electric field effect
has been found in atomically thin graphene films [1].
These materials can be regarded as bulk systems.
On the other hand, nanographenes with controlled edge
structures have been predicted to have localized states
along the zigzag edges [2].  The presence of the edge
states have been observed by experiments of scanning
tunneling spectroscopy [3,4].  Thus, the studies of the
edge states are one of the interesting topic of the field.

Previously, the magnetic switching effect has been found
in the process of insertion and extraction of molecules
in activated carbon fibers [5].  The inserted molecules
remain in nanometer size pores, and give effective
pressure to the nanographite clusters.  The Pauli 
susceptibility decreases due to the decrease of the
magnetic moment magnitude.  We have studied the magnetism
of the nanographite using the tight binding model
including the interlayer interaction between neighboring
nanographene layers [6-8].  We have found that open shell
nature of each layer might explain the experimental
observations.  Recently, interlayer hopping interaction 
effects in stacked graphite have been investigated 
theoretically [9].  The electronic structures change 
dramatically including the massless Dirac cone and parabolic 
dispersions, depending on the layer numbers from a single 
graphene, bilayer graphene, to multi layers.

In this paper, we will study the edge-state in nanographene 
materials with zigzag edges including the inter-layer 
interactions, extending the work of a single layer [2].
We will prove that edge states exist at the energy of 
the Dirac point in the doubly stacked nanographene, 
and in the case of the infinitely-wide lower layer case. 
This property can be shown both for the A-B and A-C stackings.

\section{Edge states of one graphene layer}

In this section, we review the idea of the edge state
which appears along the zigzag line of a graphene [2].
Figure 1 shows a grapnene sheet which has one zigzag
line at the top most edge of the figure.  The graphene
extends infinitely in the left, right, and down directions.
We will constitute a wavefunction of the edge state
at the energy $E=0$ as follows.  The lattice sites
are divided in A and B sublattices.  The edge atoms
belong to the A sublattice.  The edge state has the
amplitude zero at all the sites of the B sublattice.
The wavenumber in the one dimensional direction is 
denoted as $k$, and $a$ is the lattice constant of the 
unit cell, which is the length between the neighboring edge atoms.
The condition that the amplitude becomes zero 
at the atom between $(n-1)$th and $n$th edge sites, when
the nearest neighbor hopping interaction $t$ is considered, is
\beeq
{\rm e}^{ik(n-1)a} + {\rm e}^{ikna}+x=0,
\eneq
and this gives the amplitude 
\beeq
x=[-2{\rm cos}(ka/2)] {\rm e}^{ik(n-1/2)a}.
\eneq
The similar condition, 
\beeq
{\rm e}^{ikna} + {\rm e}^{ik(n+1)a}+y=0,
\eneq 
gives 
\beeq
y=[-2{\rm cos}(ka/2)] {\rm e}^{ik(n+1/2)a}.
\eneq
The condition that the amplitude becomes zero at the
sites surrounded by the amplitudes $x$, $y$, and $z$ is 
\beeq
x+y+z=0.
\eneq  
Therefore, we obtain the amplitude at the third zigzag line 
\beeq
z=[-2{\rm cos}(ka/2)]^2 {\rm e}^{ikna}.
\eneq
Iterating this process, we can prove that the amplitude
at the $n$th unit cell along the $m$th is proportional
to $[-2{\rm cos}(ka/2)]^{m-1}$.  The absolute value 
is smaller than unity if the wavenumber satisfies
$2 \pi/3 < ka \leq \pi/2$.  Therefore, the edge state
at the energy $E=0$ extends into the bulk of the graphene
with exponentially damping amplitude with respect to the 
distance from the edge.  The edge state locates near 
the boundary of the first Brillouin zone in the 
wavenumber space.

\section{Edge states in bilayer graphene with parallel zigzag lines}

Figure 2(a) shows the A-B stacked bilayer graphene with zigzag edges
at the top of the figure.  The system extends infinitely
in the other directions.  The upper layer is shown by the
solid lines, and the lower layer (dotted line) is shift 
downward by the bond length $b$.  The number $n$ ravels the 
$n$th unit cell shown by the dashed line.  The other label $m$
indicates each zigzag line.  For the convention, the upper
layer begins with $m=0$, and the lower layer starts with $m=1$.
At the circles, two carbon atoms overlap completely, and
weak hopping interaction $t_1$ is assigned here.
The same $m$ indicates the zigzag lines in the upper
and lower layers.

We will constitute wavefunctions of the edge state.
It is assumed that the edge state has amplitudes at
the sites of the A sublattice only.  The amplitude
becomes zero in the B sublattice.  Even if the interactions
$t_1$ are present at circles, the alternation of the 
A and B sites remains in the whole system. So,
the bipartite nature remains, too.  When we look at
the condition of the edge state at the energy $E=0$
of the upper layer, we find the formula, 
\beeq
a_{n,m} {\rm e}^{-ika/2} + a_{n+1,m} {\rm e}^{ika/2} 
+ a_{n,m+1} = 0,
\eneq
for even $m$, and
\beeq
a_{n,m} {\rm e}^{-ika/2} + a_{n+1,m} {\rm e}^{ika/2} 
+ a_{n+1,m+1} = 0
\eneq
for odd $m$, where $a_{n,m}$ is the
amplitude at the $n$th unit cell of the $m$th zigzag line.
Assuming $a_{n,0}=A$, we obtain 
\beeq
a_{n,m}=A[-2{\rm cos}(ka/2)]^m.
\eneq
The similar condition of the zero energy state gives,
\beeq
b_{n,m} {\rm e}^{-ika/2} + b_{n+1,m} {\rm e}^{ika/2} + b_{n,m+1} 
+ r_1 a_{n,m} = 0
\eneq
for odd $m$, and
\beeq
b_{n,m} {\rm e}^{-ika/2} + b_{n+1,m} {\rm e}^{ika/2} + b_{n+1,m+1} 
+ r_1 a_{n+1,m} = 0
\eneq 
for even $m$, for the amplitude $b_{n,m}$ of the lower layer.
Here, $r_1=t_1/t$ is the ratio of the interlayer hopping integral
to the intralayer hoppings.  As we know the form of $a_{n,m}$,
we can solve the recurrence formula of the number series
to obtain 
\beeq
b_{n,m}=-(m-1) r_1 A [-2{\rm cos}(ka/2)]^{m-2}
+ B [-2{\rm cos}(ka/2)]^{m-1}
\eneq
for $m \geq 2$, where $b_{1,m}=B$ is assumed.
Therefore, we have found that the edge state persists
in the A-B stack case.  The magnitudes of $A$ and $B$ will 
be nearly equal $A \sim B$, so the correction by the 
interlayer interaction is of the order $r_1$ with 
respect to the intralayer term.

Figure 2(b) shows the A-C stacking case, where the lower
layer is moved into the right-down direction with a
bond length $b$. The notations are similar to those
of Fig. 2(a).  The difference is the fact that the
both zigzag edge lines begin with the same index
$m=1$.  In the A-C stacking case, it is found that
the amplitude of the edge state can be constructed
in the lower layer, first.  The condition of the 
zero energy state gives, 
\beeq
b_{n,m} {\rm e}^{-ika/2} 
+ b_{n+1,m} {\rm e}^{ika/2} + b_{n,m+1} = 0
\eneq 
for even $m$, and
\beeq
b_{n,m} {\rm e}^{-ika/2} 
+ b_{n+1,m} {\rm e}^{ika/2} + b_{n+1,m+1} = 0
\eneq 
for odd $m$, for the 
amplitude $b_{n,m}$.  This sequence is solved as
\beeq
b_{n,m}=B[-2{\rm cos}(ka/2)]^{m-1},
\eneq
where $b_{n,1}=B$.  The effects of the interlayer interactions give
the next relation for the upper layer,
\beeq
a_{n,m} {\rm e}^{-ika/2} + a_{n+1,m} {\rm e}^{ika/2} 
+ a_{n,m+1} + r_1 b_{n,m} = 0
\eneq
for odd $m$, and
\beeq
a_{n,m} {\rm e}^{-ika/2} + a_{n+1,m} {\rm e}^{ika/2} 
+ a_{n+1,m+1} + r_1 b_{n+1,m} = 0
\eneq
for even $m$.  This is solved
with the help of the previous solution $b_{n,m}$ 
to give 
\beeq
a_{n,m} = A [-2{\rm cos}(ka/2)]^{m-1} - (m-1) r_1 B [-2{\rm cos}(ka/2)]^{m-2}
\eneq
for $m \geq 2$, where $a_{n,1}=A$.
The roles of the upper and lower layers
seem to be exchanged from that of the A-B stacking
case.  The amplitude of the upper layer has
correction terms of the order $r_1$ owing to
the presence of the lower layer.

\section{Edge states in bilayer graphene with infinitely-wide lower layer}

In this section, we will consider the bilayer
graphene, where the lower layer has infinite
spatial extent.  So, the nanographene with one
zigzag edge is placed upon one infinite graphene.
We consider both of the A-B and A-C stackings.
They are shown in Figs. 3(a) and (b), respectively.
As the convention of the choice of the starting
point of the initial amplitudes, $A$ and $B$, 
is different, we would like to formulate again 
for the both cases.

Figure 3(a) shows the A-B stacking case.  As the
upper layer A-site $(n,m=1)$ interacts directly
with the lower layer site as denoted by the circles,
the starting points are indexed differently from
those of the previous section.  Here, the zigzag 
lines of the upper and lower layers in the down
direction are indexed by $m$.  And, the zigzag
lines of the lower layer in the up direction
are indexed by $l$ with $l=m=1$.  In the upper
layer, the amplitude is 
\beeq
a_{n,m}=A[-2{\rm cos}(ka/2)]^{m-1}
\eneq
with $a_{n,1}=A$.  In the lower layer, the condition of 
the zero energy state is the relations,
\beeq
b_{n,m} {\rm e}^{-ika/2} + b_{n+1,m} {\rm e}^{ika/2} 
+ b_{n,m+1} + r_1 a_{n,m} = 0
\eneq
for even $m$,
\beeq
b_{n,m} {\rm e}^{-ika/2} + b_{n+1,m} {\rm e}^{ika/2} 
+ b_{n+1,m+1} + r_1 a_{n+1,m} = 0
\eneq
for odd $m$, 
\beeq
c_{n,m} {\rm e}^{-ika/2} + c_{n+1,m} {\rm e}^{ika/2} 
+ c_{n,m+1} = 0
\eneq
for even $l$ and $l \geq 2$, and
\beeq
c_{n,m} {\rm e}^{-ika/2} + c_{n+1,m} {\rm e}^{ika/2} 
+ c_{n+1,m+1} = 0
\eneq
for odd $l$.  We obtain,
\beeq
b_{n,m}=- m r_1 A [-2{\rm cos}(ka/2)]^{m-1}
+ B [-2{\rm cos}(ka/2)]^{m-1}
\eneq
and
\beeq
c_{n,l}=B[-2{\rm cos}(ka/2)]^{l-1}
\eneq
for $l \geq 2$, where $b_{n,1}=c_{n,1}=B$.

Figure 3(b) is the A-C stacking case.  In the lower layer,
the amplitude of the edge state is calculated as 
\beeq
b_{n,m}=B[-2{\rm cos}(ka/2)]^m
\eneq
and 
\beeq
c_{n,l}=B[-2{\rm cos}(ka/2)]^l,
\eneq
where $b_{n,0}=c_{n,0}=B$.  In the upper layer,
the effect of the interaction appears.
The amplitude can be derived by the relation
\beeq
a_{n,m} {\rm e}^{-ika/2} + a_{n+1,m} {\rm e}^{ika/2} 
+ a_{n,m+1} + r_1 b_{n,m} = 0
\eneq
for odd $m$ and $m \geq 1$, and
\beeq
a_{n,m} {\rm e}^{-ika/2} + a_{n+1,m} {\rm e}^{ika/2} 
+ a_{n+1,m+1} + r_1 b_{n+1,m} = 0
\eneq
for even $m$.  This is solved with the help of above 
solution $b_{n,m}$ to give 
\beeq
a_{n,m} = A [-2{\rm cos}(ka/2)]^{m-1} 
- m r_1 B [-2{\rm cos}(ka/2)]^m
\eneq
for $m \geq 2$, where $a_{n,1}=A$.
Therefore, we have constructed the wavefunction
of the zero energy state for the infinitely-wide
lower layer cases.

\section{Summary}

We have studied the edge-state in nanographene 
materials with zigzag edges including the inter-layer 
hopping interactions.  We have shown that edge states 
are present at the energy of the Dirac point 
in the doubly stacked nanographene, and in the 
case of the infinitely-wide lower layer case. 
This property has been found both for the A-B and 
A-C stackings.

\pagebreak
\begin{flushleft}
{\bf References}
\end{flushleft}

\noindent
$[1]$ K. S. Novoselov, A. K. Geim, S. V. Morozov, D. Jiang, 
Y. Zhang, S. V. Dubonos, I. V. Grigorieva, and A. A. Firsov,
Science {\bf 306}, 666 (2004).\\
$[2]$ M. Fujita, K. Wakabayashi, K. Nakada, and K. Kusakabe,
J. Phys. Soc. Jpn. {\bf 65}, 1920 (1996).\\
$[3]$ Y. Kobayashi, K. Fukui, T. Enoki, and K. Kusakabe,
Phys. Rev. B {\bf 73}, 125415 (2006).\\
$[4]$ Y. Niimi, T. Matsui, H. Kambara, K. Tagami, M. Tsukada, 
and H. Fukuyama, Phys. Rev. B {\bf 73}, 085421 (2006).\\
$[5]$ H. Sato, N. Kawatsu, T. Enoki, M. Endo, R. Kobori, 
S. Maruyama, and K. Kaneko, Solid State Commun. {\bf 125}, 641 (2003).\\
$[6]$ K. Harigaya, J. Phys.: Condens. Matter {\bf 13}, 1295 (2001).\\
$[7]$ K. Harigaya, Chem. Phys. Lett. {\bf 340}, 123 (2001).\\
$[8]$ K. Harigaya and T. Enoki, Chem. Phys. Lett. {\bf 351}, 128 (2002).\\ 
$[9]$ M. Koshino and T. Ando, Phys. Rev. B {\bf73}, 245403 (2006).\\

\pagebreak

\begin{flushleft}
{\bf Figure Captions}
\end{flushleft}

\mbox{}

\noindent
Fig. 1.  The edge states along the zigzag edge of a graphene.
See the text for the notations.

\mbox{}

\noindent
Fig. 2.  Bilayer graphene with parallel zigzag lines.
(a) A-B stacking and (b) A-C stacking are shown.
See the text for the notations.

\mbox{}

\noindent
Fig. 3.  Bilayer graphene with the infinitely-wide lower 
layer. (a) A-B stacking and (b) A-C stacking are shown.
See the text for the notations.

\pagebreak

\input epsf.tex
\epsfbox{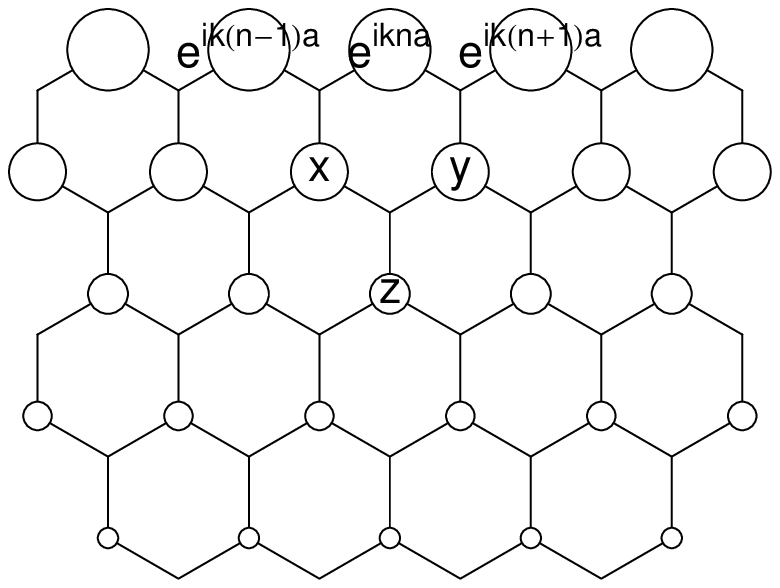}

\noindent
Fig. 1.  The edge states along the zigzag edge of a graphene.
See the text for the notations.

\pagebreak
\epsfbox{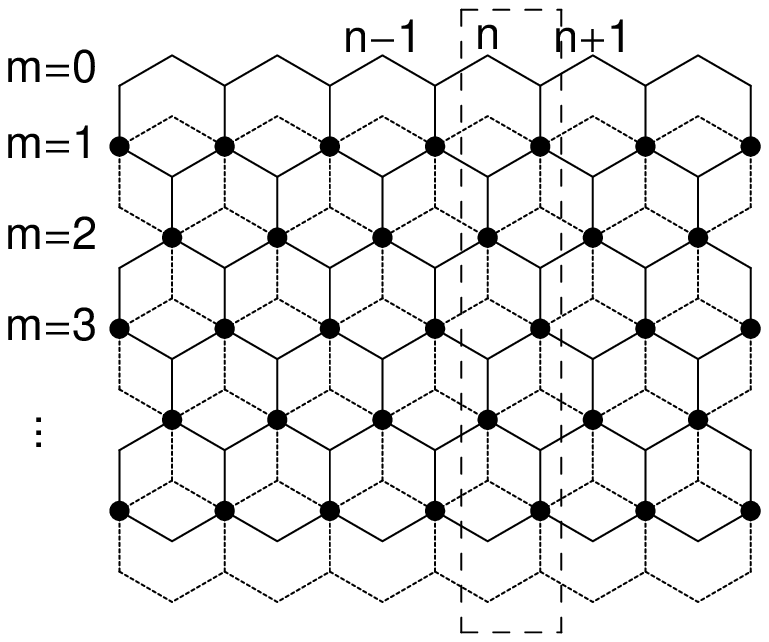}

\epsfbox{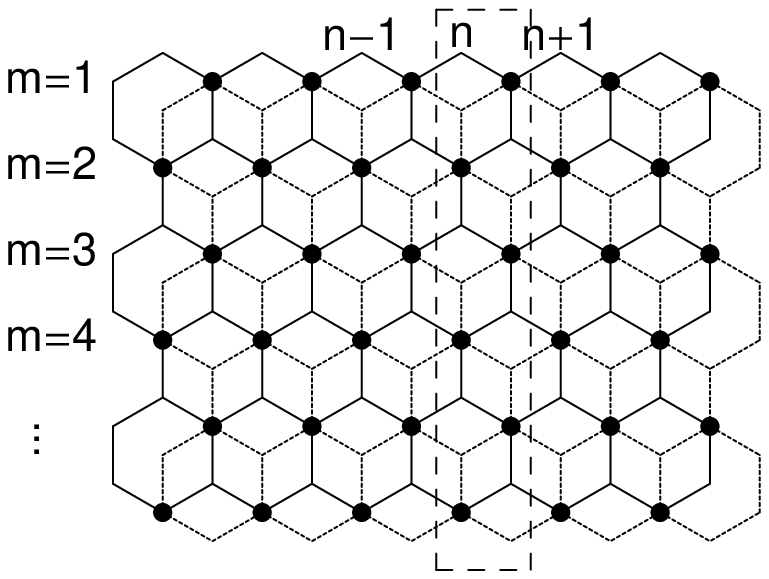}

\noindent
Fig. 2.  Bilayer graphene with parallel zigzag lines.
(a) A-B stacking and (b) A-C stacking are shown.
See the text for the notations.

\pagebreak
\epsfbox{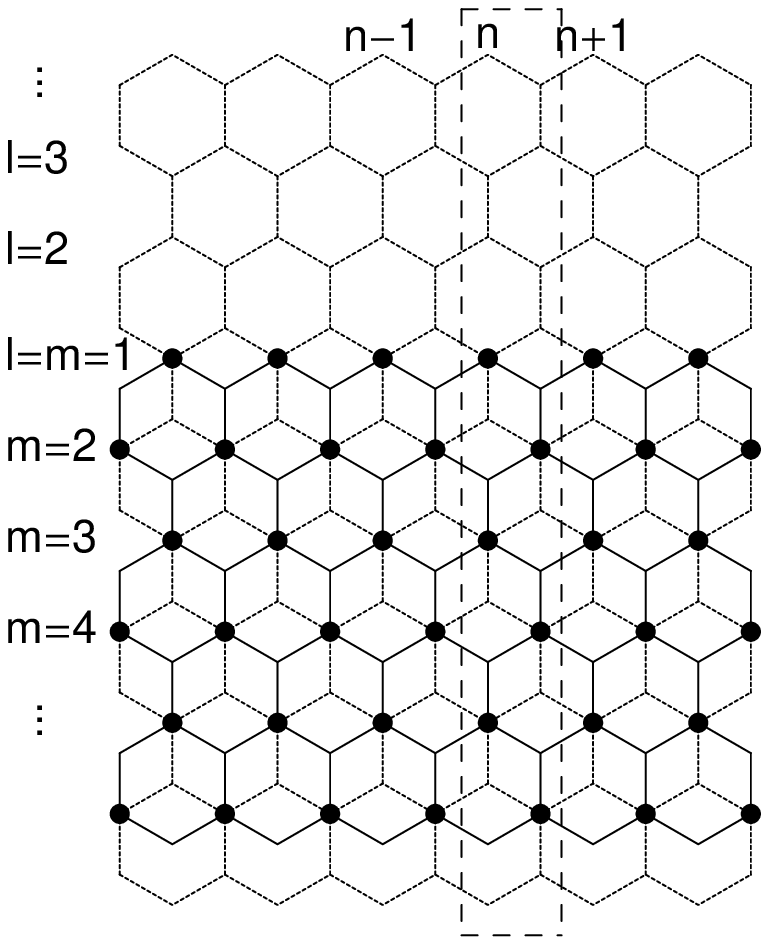}

\epsfbox{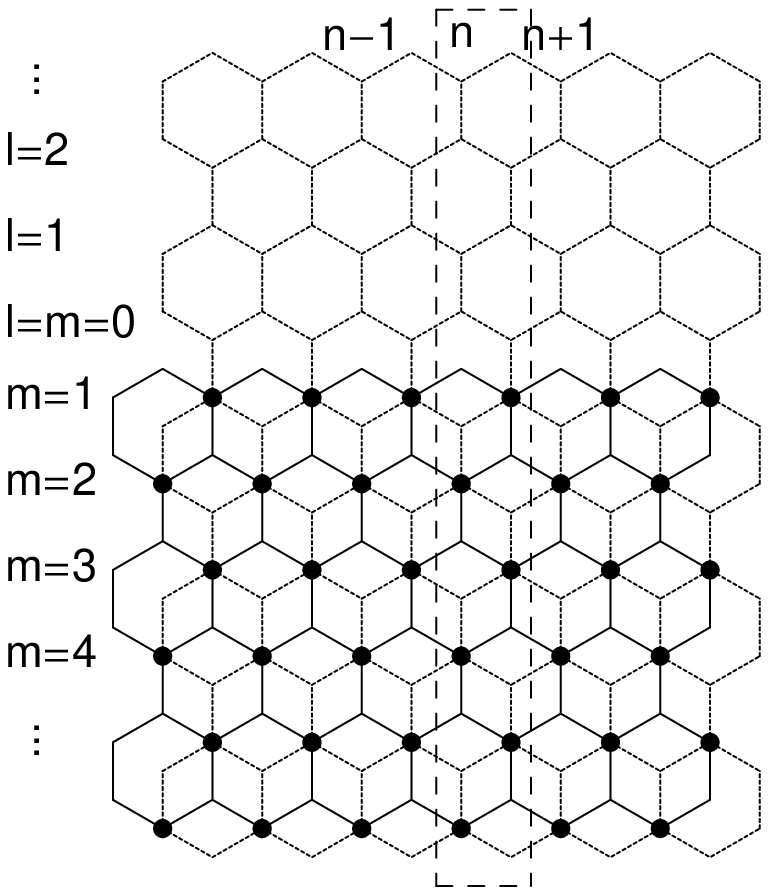}

\noindent
Fig. 3.  Bilayer graphene with the infinitely-wide lower 
layer. (a) A-B stacking and (b) A-C stacking are shown.
See the text for the notations.

\end{document}